\documentclass[aps,pra,reprint,superscriptaddress,amssymb,amsmath]{revtex4-1}
\usepackage{graphicx}
\usepackage{multirow}
\usepackage{bm}
\usepackage{mathrsfs}
\usepackage{setspace}
\usepackage{comment}
\usepackage{float}

\begin{document}

\title{Security Analysis of Measurement-device-independent Quantum Secure Direct Communication}

\author{Peng-Hao Niu}
\affiliation{State Key Laboratory of Low-Dimensional Quantum Physics and Department of Physics, Tsinghua University, Beijing 100084, China}
\affiliation{Frontier Science Center for Quantum Information, Beijing 100084, China}

\author{Jia-Wei Wu}
\affiliation{State Key Laboratory of Low-Dimensional Quantum Physics and Department of Physics, Tsinghua University, Beijing 100084, China}
\affiliation{Frontier Science Center for Quantum Information, Beijing 100084, China}

\author{Liu-Guo Yin}
\thanks{Corresponding author: yinlg@tsinghua.edu.cn}
\affiliation{Frontier Science Center for Quantum Information, Beijing 100084, China}
\affiliation{Beijing National Research Center for Information Science and Technology, Beijing 100084, China}
\affiliation{School of Information Science and Technology, Tsinghua University, Beijing 100084, China}
\affiliation{Beijing Academy of Quantum Information Sciences, Beijing 100193, China}

\author{Gui-Lu Long}
\thanks{Corresponding author: gllong@tsinghua.edu.cn}
\affiliation{State Key Laboratory of Low-Dimensional Quantum Physics and Department of Physics, Tsinghua University, Beijing 100084, China}
\affiliation{Frontier Science Center for Quantum Information, Beijing 100084, China}
\affiliation{Beijing National Research Center for Information Science and Technology, Beijing 100084, China}
\affiliation{School of Information Science and Technology, Tsinghua University, Beijing 100084, China}
\affiliation{Beijing Academy of Quantum Information Sciences, Beijing 100193, China}

\begin{abstract}

Quantum secure direct communication (QSDC) is an important branch of quantum communication that transmits confidential message directly in a quantum channel without utilizing encryption and decryption. It not only prevents eavesdropping during transmission, but also eliminates the security loophole associated with key storage and management. Recently measurement-device-independent (MDI) QSDC protocols in which the measurement is performed by an untrusted party using imperfect measurement devices have been constructed, and MDI-QSDC eliminates the security loopholes originating from the imperfections in measurement devices so that enable application of QSDC with current technology. In this paper, we complete the quantitative security analysis of the MDI-QSDC. The security capacity is derived, and its lower bound is given. It is found that the MDI-QSDC secrecy capacity is only slightly lower than that of QSDC utilizing perfect measurement devices. Therefore QSDC is possible with current measurement devices by sacrificing a small amount in the capacity.


\end{abstract}

\maketitle

\section{Introduction} 

Quantum principle provides powerful and novel technique for confidential communication~\cite{Bennett:1984wva,Ekert:1991kl,Long:2002tw} and data protection~\cite{Bennett:1983tq,Sun:2014wy}. As a unique manner of confidential communication, quantum secure direct communication (QSDC) transmits information directly through a quantum channel without establishing a secret key in advance, using entanglement~\cite{Long:2002tw,Deng:2003ih,Wang:2005tm,Wang:2006vo,Wang:2005vx,Chen:2018uo}, or single photons~\cite{Deng:2004kg}. It is in striking contrast to quantum key distribution (QKD) where only random numbers are transmitted, and information is transmitted in a subsequent classical communication in the form of ciphertext. QSDC has several distinct features, that is, no key distribution, no key management and no ciphertext, which provides additional security protection, in addition to the security in transmission, for instance, there is no threat of key loss. It has a wide range of applications, such as direct communication, authentication, quantum bidding, secret sharing and distribution of pre-determined key. Of course it can also be used to perform the key agreement just like QKD. QSDC is also source saving, for instance, in the two-step QSDC protocol~\cite{Deng:2003ih}, only one qubit transmission is required to transmit one bit of information, whereas in BB84 QKD, it requires one qubit and two classical bits where one classical bit is for comparing basis and another is for transmitting ciphertext. QSDC has been developed greatly in the last two decades, and many protocols have been proposed and
studied~\cite{Yan:2004vw,Liu:2016ve,Cao:2015vn,Wang:2018wl,Jin:2011wv,Xu:2019vt,Bebrov:2019te,Wang:2019tv,Zhukov:2018ws,Murakami:2007xx,Lu:2019uh,srikara2020ct}. Proof-of-principle experimental demonstrations have been completed in recent years. Single-photon-based QSDC protocol, DL04~\cite{Deng:2004kg}, merged with frequency coding, is demonstrated experimentally in a noisy and lossy channel~\cite{Hu:2016gz}. Demonstration of entanglement-based QSDC are completed with atomic quantum memory~\cite{Zhang:2017ew}, and optical fiber entanglement devices~\cite{Zhu:2017ie}. QSDC with data locking~\cite{Lum:2016kn}, with quantum low probability of interception~\cite{Shapiro:2019fd} and two-way QSDC with single photons~\cite{Del:2018two,Massa:2019bz} have also been proposed and demonstrated experimentally. A fully functional QSDC prototype has been built that sends secure information over 1.5km with 50bps last year~\cite{Qi:2019bi}, and 10km with 4kbps quite recently~\cite{Caict}. Quantum-memory-free QSDC protocol has also been proposed~\cite{Sun:2018vk}, which has solved one big obstacle for the practical application of QSDC.

Security is crucial in practical QSDC. Defects in devices, especially measurement devices, may lead to loopholes in practical QSDC systems, and give chances to adversaries to eavesdrop secret information~\cite{Huang:2018ts,makarov2006effects,Lydersen:2010vw}. To solve these problems, measurement-device-independent (MDI) QSDC protocols~\cite{Niu:2018cq,zhou2020measurement,Gao:2019ty} and the device-independent QSDC protocol~\cite{Zhou:2019wa} have been designed in the last two years. Specifically, device-independent QSDC can achieve secure communication even there are defects in all devices in a QSDC system. MDI-QSDC can eliminate the security loopholes associated with the measurement devices, which are the major security loopholes in practice. It introduces an untrusted third party to perform all the measurements using imperfect measurement devices, yet Alice and Bob can still ensure the security. Thus, MDI-QSDC is very appealing with the current technology. 

Security analysis of QSDC is very different from that of QKD~\cite{Kiktenko:2017ue}. In QSDC, Eve must either gets all parts of the entangled states simultaneously in entanglement-based QSDC protocols, or the states before and after the information encoding in single-photon-based QSDC protocols to steal information~\cite{Lum:2016kn}. However, Eve will be detected in the transmission of the first part of entangled pairs, or the single states before a message is encoded. Therefore Eve obtains nothing about the confidential message. Security analysis of QSDC without MDI~\cite{Qi:2019bi,Wu:2019wz} has been studied recently. It is vital to study the security of MDI-QSDC protocols and give the quantitative parameters of the protocols. In this work, we analyze the security of recently proposed MDI-QSDC protocols~\cite{Niu:2018cq,zhou2020measurement}, namely, the protocol with Einstein-Podolsky-Rosen (EPR) pairs \cite{Niu:2018cq},which we call MDI-TS QSDC, and the protocol with single photons~\cite{zhou2020measurement}, which we call MDI-DL04 QSDC hereafter. This paper is organized as follows. In section 2, we construct first an equivalent protocol of the original MDI-TS protocol~\cite{Niu:2018cq} to facilitate the analysis. The equivalent protocol does not change the security and functionality of the original MDI-TS QSDC, but makes it easier for the security analysis. In section 3, we prove the security of MDI-TS QSDC and derive the lower bound of secrecy capacity. In section 4, security of MDI-DL04 QSDC is proved, and the lower bound of its secrecy capacity is also provided. Finally, we summarize in section 5.

\section{An equivalent protocol of MDI-TS protocol}
In the MDI-TS QSDC, there are three parties, two legitimate users, Alice and Bob, and one untrusted third party, Charlie who performs all the measurements. The protocol contains 6 steps:

Step 1. Alice prepares $n$ EPR photon pairs, and takes one photon from each pairs to form a sequence, $S_A$. Alice inserts $m$ single photons at random positions of the remaining  photons, the partner photons of $S_A$, to form another sequence, $C_A$, which contains $n+m$ photons. Bob does the same and prepares two sequences,$S_B$, which contains $n$ photons from $n$ entangled photon pairs, and  $C_B$, which contains $n+m$ photons composed of $n$ photons from $n$ entangled photon pairs  and $m$ single photons inserted at random positions. The entangled photon pairs are at the Bell state $| \psi^- \rangle$, and the single photons are randomly at one of the four states, $| 0 \rangle$, $| 1 \rangle$, $| + \rangle = (| 0 \rangle + | 1 \rangle)/\sqrt{2}$ and $| - \rangle = (| 0 \rangle - | 1 \rangle)/\sqrt{2}$. The Bell states have the form shown in equation (\ref{eq:bell}).
\begin{equation}
\begin{split}
| \phi^+ \rangle &= \left(|00\rangle + |11\rangle \right) \slash\sqrt{2}, \\
| \phi^- \rangle &=   \left(|00\rangle - |11\rangle \right)\slash \sqrt{2},  \\
| \psi^+ \rangle &=   \left(|01\rangle + |10\rangle \right)\slash \sqrt{2}, \\
| \psi^- \rangle &= \left(|01\rangle - |10\rangle \right)\slash \sqrt{2}.\label{eq:bell}
\end{split}
\end{equation}

Step 2. Alice and Bob send $C_A$ and $C_B$ to Charlie respectively while keeping $S_A$ and $S_B$ in their hands.

Step 3. Charlie makes Bell measurement on photons pairwise, one from $C_A$ and one from $C_B$, then announces the results to Alice and Bob. According to the photons provided by $C_A$ and $C_B$, there will be three cases: (i), both photons are partner photons from EPR pairs. (ii), both photons are single photons, and (iii), one photon is a single photon and the other is a partner photon from an EPR pair. For these three cases, case (iii) will be discarded for simplicity and symmetry, while case (ii) will be used for security check, and case (i) is utilized to transmit secret message because a photon in $S_A$ and the corresponding photon in $S_B$  will be entangled due to entanglement swapping.

Step 4. Security check. After receiving the results of Bell measurement, Alice and Bob exchange position information of their $m$ single photons and the corresponding basis information. For single photons prepared in the same basis, the decomposition in terms of Bell states is shown in equation (\ref{eq:dos})
\begin{equation}
\begin{split}
|\; 0 \; 0 \;\rangle &= \left(|\phi^+\rangle + |\phi^-\rangle \right)/\sqrt{2}, \\
|\; 1 \; 1 \;\rangle &= \left(|\phi^+\rangle - |\phi^-\rangle \right)/\sqrt{2}, \\
|\; 0 \; 1 \;\rangle &= \left(|\psi^+\rangle + |\psi^-\rangle \right)/\sqrt{2}, \\
|\; 1 \; 0 \;\rangle &= \left(|\psi^+\rangle - |\psi^-\rangle \right)/\sqrt{2}, \\
|+  + \rangle &= \left(|\phi^+\rangle + |\psi^+\rangle \right)/\sqrt{2}, \\
|-  - \rangle &= \left(|\phi^+\rangle - |\psi^+\rangle \right)/\sqrt{2}, \\
|+  - \rangle &= \left(|\phi^-\rangle - |\psi^-\rangle \right)/\sqrt{2}, \\
|-  + \rangle &= \left(|\phi^-\rangle + |\psi^-\rangle \right)/\sqrt{2}.
\label{eq:dos}
\end{split}	
\end{equation}

An adversary's attack will be identified because her action will cause inconsistent results in the Bell measurement, and Charlie's dishonest behavior will also be found.

Step 5. Encoding the message. After confirming the security and  removing redundant photons used in the above steps, the particles shared by Alice and Bob are now all in Bell states, and we name the particle sequence at Alice and Bob as $M_A$ and $M_B$, respectively. Alice will use four kinds of unitary operations to encode messages, namely $U_{00} = I, U_{01} = \sigma_x, U_{10} = i\sigma_y, U_{11} = \sigma_z$ to represent $00,01,10,11$ respectively, which are exactly the dense coding operations. Meanwhile, Bob performs $U_{i,j}, i,j \in\{0,1\}$ randomly on photons in $M_B$. We call this operation as a ``cover" operation, because after this operation, states of each photon pair in $M_A$ and $M_B$ will be completely random in four kinds of Bell states and the details are only known to Bob.

Step 6.  Alice and Bob send the encoded $M_A$ and $M_B$ sequences to Charlie, and Charlie performs the Bell measurement. Charlie then publicly announces the measurement results, and Bob can decode the message by combining the measurement results and initial states of $M_A$ and $M_B$.

We use the idea of virtual qubits~\cite{Lo:2012dfa} to construct the equivalent protocol. In Step 1, the inserted single photons are replaced by entangled photons when preparing $C_A$ and $C_B$. This means, $S_A$ and $C_A$ (also $S_B$ and $C_B$) have the same sequence length with all particles in Bell states. Then for security check, Alice or Bob performs corresponding unitary operations according to Charlie's measurement results to complete entanglement swapping, which changes Bell states shared by Alice and Bob into a uniform state, such as the singlet state, in order to simplify the parameter estimation. Alice and Bob will apply local measurements in the $Z$ or $X$ basis, on photons in $S_A$ and $S_B$ in their hands respectively. Security check will be carried out by Alice and Bob who announce the measurement results and compare them if the results are the same or not. Quantum bit error rates (QBERs), such as bit error rate and phase error rate, are also obtained from these measurement results. Consequently, the equivalent MDI-TS protocol contains the following steps:

Step $1'$. Alice prepares $n$ EPR photon pairs at Bell state $| \psi^- \rangle$, and splits each entangled pair into two parts. One part forms sequence $S_A$ and another part is $C_A$. Bob will do the same and prepares $S_B$ and $C_B$.

Step $2'$. Alice and Bob send $C_A$ and $C_B$ to Charlie for Bell measurement while keeping $S_A$ and $S_B$ in their hands respectively.

Step $3'$. Charlie announces measurement results, and Alice or Bob performs proper local unitary operations to complete the entanglement swapping. Then Alice and Bob make random local measurements in basis $Z$ or $X$ on randomly selected photons from $S_A$ and $S_B$. Photons of $S_A$ and $S_B$ are entangled and the measurement results will be correlated parallel or anti-parallel if both photons of an EPR pair are measured under the same basis, which is $XX$ or $ZZ$.

Step $4'$. After ensuring the security, Alice and Bob discard photons whose partner photons have been used in the preceding security check and obtain $M_A$ and $M_B$. Then Alice encodes the message using dense coding unitary operations $U_{ij}, (i,j \in \{0,1\})$, which were defined in Step 5, and Bob will apply an arbitrary unitary operation $U_{i,j}$ as cover operation randomly on each photon in $M_B$. 

Step $5'$. Alice and Bob send the encoded $M_A$ and $M_B$ sequences to Charlie for the Bell measurement. Bob then decodes the message from Alice after Charlie announces the measurement results.

This modified MDI-TS QSDC is an equivalent transformation of the original one, and is convenient for security analysis.

\section{Security analysis of the equivalent MDI-TS protocol}

In MDI-QSDC, confidential message is transmitted between legal users, and the potential adversary is kept ignorant of the content. There are also broadcast channels between the three participants for the necessary classical information to execute the protocol. We denote the system of Alice, Bob and the adversary Eve as $A$, $B$ and $E$ respectively. Then the secrecy capacity $C_S$ between two legal participants is given as (\ref{eq:cs}) according to $\mathrm{Csisz\acute{a}r} - \mathrm{K\ddot{o}rner}$ theory \cite{Csiszar:1978tx},
\begin{equation}
	C_S = \mathrm{max} \left[I(A:B) - I(A:E) \right]\label{eq:cs}.
\end{equation}
where the exist of  $C_S$ implies there is a forward encoding scheme with lower capacity than $C_S$, it can be used to transmit the message reliably and securely to receivers. This is quite different from QKD which relies on post-processing. 

In the equivalent MDI-TS protocol, we are ignorant of the measurement process and  strategy that an adversary may exploit, hence we focus on the system after the entanglement swapping, where a joint state $\rho_{AB}^{jnt}$, consisting of photon pairs shared between Alice and Bob. We consider a situation where Eve attacks the system with an auxiliary system $| E \rangle$ and performs coherent attack. According to quantum De Finetti theorem \cite{Renner:2007bc}, we can use a direct product of independent and identically distributed (i.i.d.) subsystems $\rho_{AB}^{\otimes n}$ to approximate $\rho_{AB}^{jnt}$ asymptotically, if a randomized permutation is applied to the system, that is,  Alice and Bob can permute the particles in their hands with $\mathcal{P}_\mu \rho_{AB}^{jnt} \mathcal{P}^{\dagger}_\mu$, where $\mathcal{P}_{\mu}$ is the permutation operators and the same permutation is chosen with some classical information $\mu$. In this case, Eve's attacks can be considered as collective attack, and it is a sufficient deliberation that $|\Phi_{ABE}\rangle$ is a purification of $\rho_{AB}$. Using the method in \cite{Kraus:2005ki}, we can erase non-diagonal elements of $\rho_{AB}$ in the Bell basis with local unitary operations $\mathcal{U}_\nu$ and construct $\rho_{AB}$ in bases $\{|\psi^-\rangle, |\psi^+\rangle, |\phi^-\rangle, |\phi^+\rangle\}$ as

\begin{equation}
	\begin{split}
		\rho_{AB} =\, &\delta_1 |\psi^-\rangle \langle \psi^-| + \delta_2 |\psi^+\rangle \langle \psi^+| \\
		+ &\delta_3 |\phi^-\rangle \langle \phi^-| + \delta_4 |\phi^+\rangle \langle \phi^+|,
	\end{split}
\end{equation}
where $\sum\limits_{i=1}^4 \delta_i = 1$. Then $|\Phi_{ABE}\rangle$ is
\begin{equation}
	|\Phi_{ABE} \rangle = \sum_i^4 \sqrt{\delta_i} |\Psi_i \rangle |E_i\rangle,
\end{equation}
where $| \Psi_i \rangle$ is the Bell states shared by Alice and Bob, and $|E_i\rangle$ is the orthogonal states of system $| E \rangle$.

In security check of Step $3'$, the bit error $\epsilon_z$ and phase error $\epsilon_x$ will constrain parameters $\delta_i$ as $\epsilon_z = \delta_3 + \delta_4$ and $\epsilon_x  = \delta_2 + \delta_4$. Alice and Bob will then carry out the message encoding process in Step $4'$. Bob performs cover operations, and we have
\begin{equation}
	\begin{split}
	\rho_{ABE}^c & = \frac{1}{4} (  | \Phi_{ABE} \rangle \langle \Phi_{ABE} |+ \sigma_{x}^B | \Phi_{ABE} \rangle \langle \Phi_{ABE} | \sigma_{x}^B \\
	 & + \sigma_{y}^B  | \Phi_{ABE} \rangle \langle \Phi_{ABE} | \sigma_{y}^B +\sigma_{z}^B | \Phi_{ABE} \rangle \langle \Phi_{ABE} | \sigma_{z}^B ).
	\end{split}
\end{equation}
After Alice's encoding operation, it becomes
\begin{equation}
	\rho_{ABE}^{ij} = U_{ij}\rho_{ABE}^c U^{\dagger}_{ij},
\end{equation}
where $i,j \in \{0,1\}$ which represents the four unitary operations described in Step $4'$.

We denote the codeword that Alice encodes as $\mathcal{C} = \{\zeta_1,\dots,\zeta_m\}$, where $\zeta_i \in \{00,01,10,11\}, i = 1,2,\dots,m$ and stands for a two-bits symbol, and the probability of $\mathcal{C}$ is $p_{\mathcal{C}}$. Then employing Holevo bound \cite{Holevo:1973wy}, we can calculate the mutual information $I(\mathcal{A}:\mathcal{E})$ of the joint system $AE$, supposing that each symbol in $\{\zeta_i\}$ has the same distribution $p_{\zeta_i} = \frac{1}{4}$. Thus, we have
\begin{equation}
	\begin{split}
	I(\mathcal{A}:\mathcal{E}) &\leq S\left(\sum_{{\mathcal{C}}}p_{\mathcal{C}}\rho_{ABE}^{\mathcal{C}}\right) - \sum_{{\mathcal{C}}}p_{\mathcal{C}}S\left(\rho_{ABE}^{\mathcal{C}}\right) \\
	&\leq m \left[S\left( \sum_{{\zeta}} p_{\zeta} \rho_{ABE}^{\zeta}  \right) - 2 \right],
	\end{split}
\end{equation}
where $S(\cdot)$ is the von Neumann entropy and $\zeta$ represents $\zeta_i$ for succinctness. After some calculation, we can obtain the subsystem mutual information of $AE$
\begin{equation}
	I(A:E) \leq h(\epsilon_z) + h(\epsilon_x),
\end{equation}
where $h(\cdot)$ is the binary entropy function. We denote the gain of Bob for message decoding as $Q$ and the gain gap between the channels of $AB$ and $AE$ as $\eta$. The error rates obtained from message decoding are noted as $\mathscr{E}$, which is a vector with four components (remember $\zeta \in \{00,01,10,11\}$) representing the error rate distribution. If the main channel between Alice and Bob is a symmetric discrete one, and the input symbol with equal probability, the secrecy capacity is 
\begin{equation}
	C_S \geq Q\left\{2-H(\mathscr{E}) - \eta \left[h(\epsilon_z) + h(\epsilon_x)\right]\right\} \label{eq:sc},
\end{equation}
according to equation (\ref{eq:cs}), where $H(\cdot)$ is the Shannon entropy.

It should be noted that we have utilized permutation operations $\mathcal{P}_{\mu}$ and local unitary operations $\mathcal{U}_\nu$ to simplify the analysis. These two operations both employ the public channel to exchange classical information $\mu$ and $\nu$, which may cause additional information leakage. Hence equation (\ref{eq:sc}) serves as the lower bound of secrecy capacity if we remove the two operations. 

\begin{figure}[htbp]
\centering
\includegraphics[width = 0.5\textwidth]{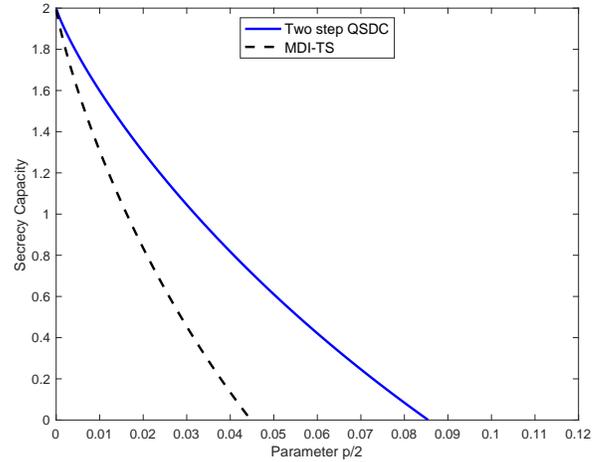}
\caption{Secrecy capacity of MDI-TS QSDC over quantum channel parameter $p/2$.}\label{mdi-qsdc01}
\end{figure}

As an example, we apply the above analysis to the depolarizing channel where the change of the channel is $\rho \Rightarrow \hat{\rho} = p \frac{\mathbb{I}}{2} + (1-p) \rho $, and $p$ is a parameter describing the channel, and $Q=\eta =1$. We perform numerical simulation, and the simulation results are shown in Fig.\ref{mdi-qsdc01}. The channel parameter of MDI-TS protocol describes quantum channel of one side, and we assume that a symmetric channel model, which implies that the two channels, Alice to Charlie and Bob to Charlie, of MDI-TS protocol are totally the same. We can find that compared with two-step QSDC, MDI-TS QSDC have lower secrecy capacity. The reason is apparent because there are two quantum channels utilized in the MDI-TS QSDC, hence more influences from the depolarizing effect of the channel.

\section{Security analysis of MDI-DL04 Protocol}

DL04 protocol uses single photons for QSDC. We now analyze the security of MDI-DL04 QSDC. Suppose Alice is going to send secret message to Bob. Alice and Bob prepare photon sequences $S_A$ and $S_B$ respectively. $S_A$ consists of entangled photons and single photons as in the original MDI-TS protocol, while $S_B$ contains only single photons. Single photons in both $S_A$ and $S_B$ are randomly prepared in nonorthogonal bases. Single photons in $S_A$ will be used for security check, while entangled photon pairs will be used for message transmission. Now we give the equivalent MDI-DL04 protocol first. Step $1''$ to Step $3''$ in the equivalent protocol are the same as Step $1'$ to Step $3'$, and 

Step $4''$. After ensuring the security, Alice and Bob discard redundant photons used in previous steps, and end up with $M_A$ and $M_B$. Alice encodes message with unitary operations $U_0 = I$ and $U_1 = \sigma_u$, which represent classical bit $0$ and $1$ respectively. The operation $\sigma_u$ is one of the three unitary operations $\{\sigma_x, i\sigma_y, \sigma_z\}$. Alice chooses one of them to represent classical bit $1$ and keeps using it until one round communication is completed,  and she tells Bob what operation she has chosen. 

Step $5''$. Alice sends $M_A$ to Charlie for single photon measurement on each of the photons. Bob tells Charlie the measurement basis to be used, which depends on $U_1$. If Alice has chosen $U_1 = \sigma_x$ then Charlie should use basis $Z$, or basis $X$ if Alice has chosen $U_1 = \sigma_z$. Charlie gets measurement results $R_{M_A}$ and announces it to Bob. While photons in $M_B$, which is still held by Bob and were entangled with those in $M_A$ after Charlie's Bell measurement, will collapse into single photon states. Bob will use the same basis that Charlie used and make single photon measurement, and obtain results $R_{M_B}$. Combining with $R_{M_A}$, Bob can decode the message that Alice has encoded. It should note that if $U_1 = i\sigma_y$ is used in Step $4''$, then basis $Y$ should also be used in Step $3''$ in order to obtain error rate $\epsilon_y$ under basis $Y$. 

\begin{figure}[H]
\centering
\includegraphics[width = 0.5\textwidth]{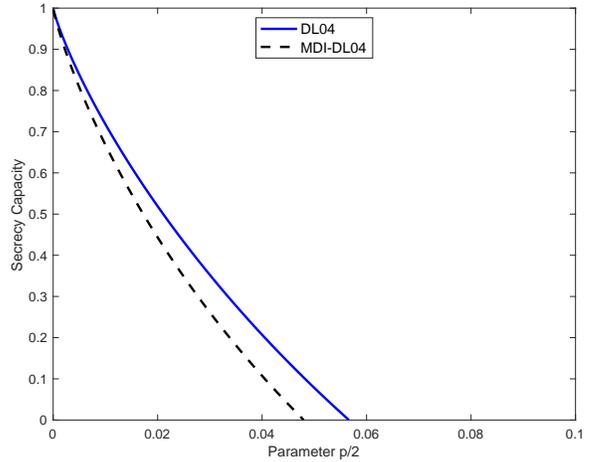}
\caption{Secrecy capacity of MDI-DL04 QSDC over quantum channel parameter $p/2$.}\label{mdi-dl04}
\end{figure}

Now we analyze the security of MDI-DL04 QSDC. Comparing the equivalent protocol of original MDI-TS QSDC (Step $1'-5'$) and MDI-DL04 QSDC (Step $1''-5''$), differences are only in the encoding and decoding process of the message. Bob keeps $M_B$ in hand without sending out, hence the state of system $AE$ is
\begin{equation}
	\rho_{AE}^{k} = U_{k}\rho_{AE}U^{\dagger}_{k},
\end{equation}
where $k \in \{0,1\}$, $\rho_{AE} = \mathrm{Tr}_B \left( | \Phi_{ABE}\rangle \langle \Phi_{ABE} | \right)$. Consequently, the secrecy capacity satisfies
\begin{equation}
	C_s^{T} \geq Q\left[1-h(e) - \eta h(\epsilon_u)\right],
\end{equation}
where $e$ is the error rate from the message, and $\epsilon_u$ is the error rate of the basis in $U_1$ that used when encoding the message. We simulated the secrecy capacity of MDI-DL04 and the result is shown in Fig.\ref{mdi-dl04}. When estimating the QBERs, basis $Y$ is used in MDI-DL04, hence $\epsilon_{u} = \epsilon_y$, while $X$ and $Z$ are used in DL04. We can see that MDI-DL04 has a lower secrecy capacity compared with DL04 without MDI. This is expected because we have more transmissions and operations. 

However, the secrecy capacity difference between MDI-DL04 and DL04 is not as large as that in Fig.\ref{mdi-qsdc01} for entanglement QSDC protocols. This is because in DL04, both $X$ and $Z$ are used, resulting in the leakage information subject to $I(A:E)_\textrm{DL04} \leq h(\epsilon_x + \epsilon_z)$~\cite{Qi:2019bi}, indicating a lower secrecy capacity estimation of non-MDI DL04 where the constraint between error rates is $\textrm{max}(\epsilon_y) = \epsilon_x + \epsilon_z$. Relatively, in equivalent MDI-DL04 protocol, only one basis is used, therefore a lower bound is obtained~\cite{Wu:2019wz}. For that reason, the difference between the secrecy capacities of MDI-DL04 and non-MDI DL04 protocols diminishes.

\section{Summary}

In summary, we have analyzed the security of two MDI-QSDC protocols, MDI-TS and MDI-DL04, that eliminate measurement devices related loopholes. Lower bound of the secrecy capacity of these two protocols is derived. MDI-QSDC has a slightly smaller secrecy capacity than that of a non-MDI QSDC protocol due to its higher security requirement. MDI-QSDC increases the security of QSDC, enabling it with current imperfect measurement devices. As two users send their photons to a third party in the middle, the MDI-QSDC can efficiently double the communication distance. The security analysis of these protocols has fulfilled an essential requirement for the practical application of QSDC.

\begin{acknowledgments}
This work was supported by the National Basic Research Program of China (2017YFA0303700), the Key R\&D Program of Guangdong province (2018B030325002), the National Natural Science Foundation of China under Grants No. 61726801, No. 11974205, and No. 11774197, and in part by the Beijing Advanced Innovation Center for Future Chip (ICFC).
\end{acknowledgments}

\bibliographystyle{apsrev4-1}
\bibliography{sa4mdi.bib}

\end{document}